\documentclass[preprint2]{aastex61}

\newcommand\aastex{AAS\TeX}

\usepackage{epstopdf}

\received{June 1, 2017}
\revised{June 1, 2017}
\accepted{\today}

\submitjournal{ApJ}

\shorttitle{\aastex\ Simultaneous transverse and longitudinal prominence oscillations}
\shortauthors{Zhang et al.}

\begin{document}

\title{Simultaneous transverse and longitudinal oscillations in a quiescent prominence triggered by a coronal jet}

\correspondingauthor{Qingmin Zhang}
\email{zhangqm@pmo.ac.cn}

\author[0000-0003-4078-2265]{Q. M. Zhang}
\affil{Key Laboratory for Dark Matter and Space Science, Purple Mountain Observatory, CAS, Nanjing 210008, China}
\affil{CAS Key Laboratory of Solar Activity, National Astronomical Observatories, Beijing 100012, China}

\author{D. Li}
\affil{Key Laboratory for Dark Matter and Space Science, Purple Mountain Observatory, CAS, Nanjing 210008, China}

\author{Z. J. Ning}
\affil{Key Laboratory for Dark Matter and Space Science, Purple Mountain Observatory, CAS, Nanjing 210008, China}

\begin{abstract}
In this paper, we report our multiwavelength observations of the simultaneous transverse and longitudinal oscillations in a quiescent prominence. 
The prominence was observed by the Global Oscillation Network Group and by the Atmospheric Imaging Assembly on board the \textit{Solar Dynamics Observatory} (\textit{SDO}) on 2015 June 29.
A \textit{GOES} C2.4 flare took place in NOAA active region 12373, which was associated with a pair of short ribbons and a remote ribbon.
During the impulsive phase of the flare, a coronal jet spurted out of the primary flare site and propagated in the northwest direction at an apparent speed of $\sim$224 km s$^{-1}$.
Part of the jet stopped near the remote ribbon. The remaining part continued moving forward before stopping to the east of prominence. 
Once the jet encountered the prominence, it pushed the prominence to oscillate periodically. The transverse oscillation of the eastern part (EP) of prominence can be divided into 
two phases. In phase I, the initial amplitude, velocity, period, and damping timescale are $\sim$4.5 Mm, $\sim$20 km s$^{-1}$, $\sim$25 minutes, and $\sim$7.5 hr, respectively. 
The oscillation lasted for two cycles. In phase II, the initial amplitude increases to $\sim$11.3 Mm while the initial velocity halves to $\sim$10 km s$^{-1}$. The period increases by a factor of $\sim$3.5.
With a damping timescale of $\sim$4.4 hr, the oscillation lasted for about three cycles.
The western part (WP) of prominence also experienced transverse oscillation. The initial amplitude is only $\sim$2 Mm and the velocity is less than 10 km s$^{-1}$.
The period ($\sim$27 minutes) is slightly longer than that of EP in phase I. The oscillation lasted for about four cycles with the shortest damping timescale ($\sim$1.7 hr).
To the east of prominence, a handful of horizontal threads experienced longitudinal oscillation. The initial amplitude, velocity, period, and damping timescale are $\sim$52 Mm, $\sim$50 km s$^{-1}$, 
$\sim$99 minutes, and 2.5 hr, respectively. To our knowledge, this is the first report of simultaneous transverse and longitudinal prominence oscillations triggered by a coronal jet.
\end{abstract}


\section{Introduction} \label{sec:intro}
Solar prominences are cool and dense plasma structures suspending in the corona with diverse morphology and rich dynamics \citep[][and references therein]{lab10,mac10}. The densities of prominences are $\sim$100 times 
larger than the corona, while the temperatures of prominences are $\sim$100 times lower than the corona. They can be observed in radio \citep{gop03}, Ca\,{\sc ii}\,H \citep{ber08,ning09a}, H$\alpha$ \citep{eng76,hao15}, 
and extreme-ultraviolet (EUV) wavelengths \citep{hei08,mcc15}. When observed on disk, the bright prominences appear as dark filaments in the filament channels along the polarity inversion lines (PILs) \citep{van89,mar98}. 
They can be found in the quiet region (QR), active regions (ARs), and near the polar region with high latitudes \citep{ler83,su12}. Magnetohydrostatic (MHS) equilibrium condition of a filament requires that the 
gravitational force on the filament is balanced by the upward magnetic tension force of the dips, either in sheared arcades \citep{anti94,liu12,zqm15} or in twisted magnetic flux ropes \citep{mar01,kepp14,ter16}. 
Sometimes, a flux rope and a dipped arcade can coexist along one filament \citep{guo10}. Filaments are divided into normal-polarity and inverse-polarity types \citep{pri89,ou17}. 
Oscillations are excited in prominence structures when they interact with propagating disturbances such as coronal EIT waves and chromospheric Moreton waves.
The periods of oscillations range from a few to tens of minutes \citep{iso06,ning09b,sch13}, and the displacements range from a few to tens of Mm \citep{oka07,kim14}. According to the velocity amplitude, they can be 
classified into small-amplitude ($\le$3 km s$^{-1}$) and large-amplitude ($\ge$20 km s$^{-1}$) oscillations. In most cases, the amplitudes of oscillations damp with time \citep[e.g.,][]{her11,gos12}. 
Recently, a rare case of growing amplitudes of filament oscillations is reported, which is explained by the thread-thread interaction \citep{zqm17,zhou17}.
Based on the direction of oscillation with respect to the filament axis, they can be divided into transverse \citep[e.g.,][]{hyd66,ram66,kle69,chen08} 
and longitudinal oscillations \citep[e.g.,][]{jing03,vrs07,zqm12,li12,bi14,chen14,luna14,zheng17}. A prominence can even undergo transverse and longitudinal oscillations simultaneously \citep{pant16,wang16}.

The triggering mechanism of filament oscillations is a very important issue. The large-amplitude transverse oscillations are often caused by Moreton waves and EUV waves from a remote site of eruption at speeds of $\sim$1000 
km s$^{-1}$ \citep[e.g.,][]{eto02,gil08}. The strong and impulsive impact of waves can shake the filaments and trigger oscillations. The large-amplitude longitudinal oscillations, however, are usually
triggered by flares or subflares near the footpoints of filaments \citep{jing03,jing06,vrs07,li12,zqm12}. The localized plasma pressure increases impulsively during the flares or subflares, which propels the filament to oscillate 
around the magnetic dips \citep{zqm13}. The component of gravity along the dip serves as the restoring force \citep{luna12a,luna12b,luna16a,luna16b}. Once the initial amplitude exceeds a critical value, 
chances are that part of the filament material undergoes downward drainage into the chromosphere, while the remaining part continues to oscillate \citep{zqm13,zqm17}. Considering that the filaments are three-dimensional (3D) 
in nature and are often supported by magnetic flux ropes, the flares or subflares may also result in enhancement of magnetic pressure, which drives longitudinal oscillations \citep{vrs07}. Magnetic pressure gradient
is considered as the restoring force, and the poloidal magnetic field of the flux rope can be estimated. In the case of 2010 August 20, the longitudinal oscillation was triggered by episodic jets connecting the energetic event 
and the filament threads \citep{luna14}. On 2016 January 26, interaction between two filaments took place in a long filament channel. During the interaction, longitudinal filament oscillation was triggered by
the moving plasma at a speed of $\sim$165 km s$^{-1}$ from the flare region \citep{zheng17}. Sometimes, when a coronal shock wave impacts a nearby filament during its propagation, 
it can trigger both transverse and longitudinal filament oscillations \citep{shen14,pant16}. Transverse oscillations in coronal loops triggered by coronal jets have been observed \citep{sar16}. 
However, simultaneous transverse and longitudinal oscillations in a prominence triggered by a coronal jet have never been investigated.

In this paper, we report our multiwavelength observations 
of the large-amplitude oscillations of a quiescent prominence triggered by the jet from a remote C2.4 solar flare on 2015 June 29. The paper is structured as follows. Data analysis is described in Section~\ref{sec:data}. 
Results are shown in Section~\ref{sec:result}. Discussions about the triggering mechanism are arranged in Section~\ref{sec:discuss}. Finally, we give a brief summary in Section~\ref{sec:summary}.

\section{Instruments and data analysis} \label{sec:data}

Located north to NOAA AR 12373 (N15E53), the prominence was continuously observed by the Global Oscillation Network Group (GONG) in H$\alpha$ line center (6562.8 {\AA})
and by the Atmospheric Imaging Assembly \citep[AIA;][]{lem12} 
on board the \textit{Solar Dynamics Observatory} (\textit{SDO}) in UV (1600 {\AA}) and EUV (94, 171, 304, and 211 {\AA}) wavelengths. The photospheric line-of-sight (LOS) magnetograms were observed by the Helioseismic and 
Magnetic Imager \citep[HMI;][]{sch12} on board \textit{SDO}. The level\_1 data from AIA and HMI were calibrated using the standard \textit{Solar Software} (\textit{SSW}) programs \textit{aia\_prep.pro} and \textit{hmi\_prep.pro}. 
The full-disk H$\alpha$ and AIA 304 {\AA} images were coaligned with an accuracy of $\sim$1$\farcs$2 using the cross correlation method. The global coronal 3D magnetic configuration 
was derived from the potential field source surface \citep[PFSS;][]{sch03} modeling. The EUV flux in 1$-$70 {\AA} was recorded by the Extreme Ultraviolet Variability Experiment \citep[EVE;][]{wood12} on board \textit{SDO}.
The soft X-ray (SXR) flux in 1$-$8 {\AA} was recorded by the \textit{GOES} spacecraft. The observational parameters, including the instrument, wavelength, time, cadence, and pixel size are summarized in Table~\ref{tab:para}.

\begin{deluxetable*}{ccccc}
\tablecaption{Description of the observational parameters \label{tab:para}}
\tablecolumns{5}
\tablenum{1}
\tablewidth{0pt}
\tablehead{
\colhead{Instrument} &
\colhead{$\lambda$} &
\colhead{Time} & 
\colhead{Cadence} & 
\colhead{Pixel size} \\
\colhead{} & 
\colhead{({\AA})} &
\colhead{(UT)} & 
\colhead{(s)} & 
\colhead{(\arcsec)}
}
\startdata
GONG & 6562.8 & 17:00$-$20:55 & 60 & 1.0 \\
\textit{SDO}/AIA & 94, 171, 211, 304 & 17:00$-$03:00+1d & 12 & 0.6 \\
\textit{SDO}/AIA & 1600 & 17:00$-$03:00+1d & 24 & 0.6 \\
\textit{SDO}/HMI & 6173 & 17:00$-$03:00+1d & 45 & 0.5 \\
\textit{SDO}/EVE & 1$-$70 & 17:00$-$19:00 & 0.25 & \nodata \\
\textit{GOES} & 1$-$8 & 17:00$-$19:00 & 2.05 & \nodata \\
\enddata
\end{deluxetable*}

\section{Results} \label{sec:result}

\subsection{Magnetic field and configuration} \label{s-magnetic}
\begin{figure*}
\plotone{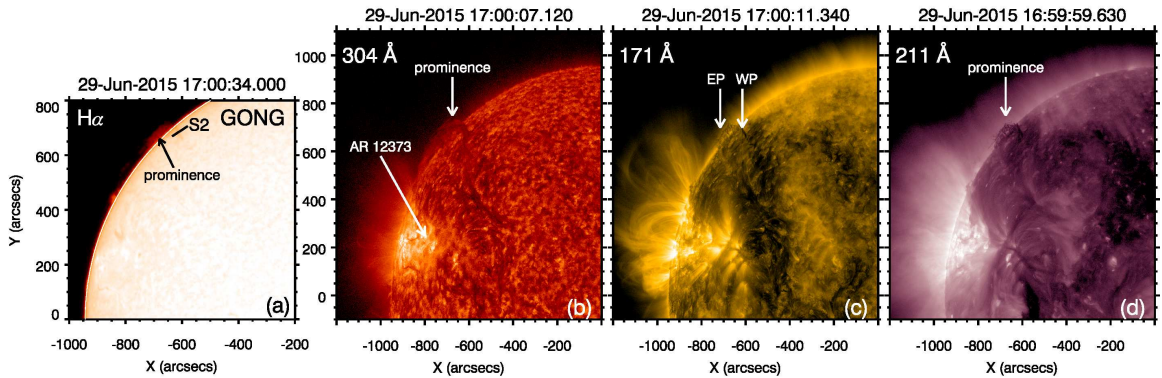}
\caption{The prominence and AR 12373 observed before the C2.4 flare in H$\alpha$, 304, 171, and 211 {\AA}, respectively.
The short slice (S2) in panel (a) is used to investigate the transverse oscillation of WP.
\label{fig1}}
\end{figure*}

Figure~\ref{fig1} shows the quiescent prominence (N45E90) and AR 12373 in various wavelengths around 17:00 UT. The prominence consisted of two parts, the eastern part (EP) and western part (WP).
The EP of prominence was composed of many bright vertical threads in H$\alpha$. 
However, it appeared as dark fine threads in EUV 171 and 211 {\AA} images due to its low temperature ($\sim$0.01 MK) compared to the ambient corona ($\sim$1 MK). 
According to recent statistical results, nearly 96\% of the quiescent filaments are associated with a flux rope magnetic configuration, while only 4\% are associated with a sheared arcade configuration \citep{ou17}. 
Thus, we believe that the vertical threads indicate a flux rope morphology of the prominence magnetic field. The WP of prominence resembles a vertical column and is shorter than EP.

\begin{figure*}
\epsscale{.80}
\plotone{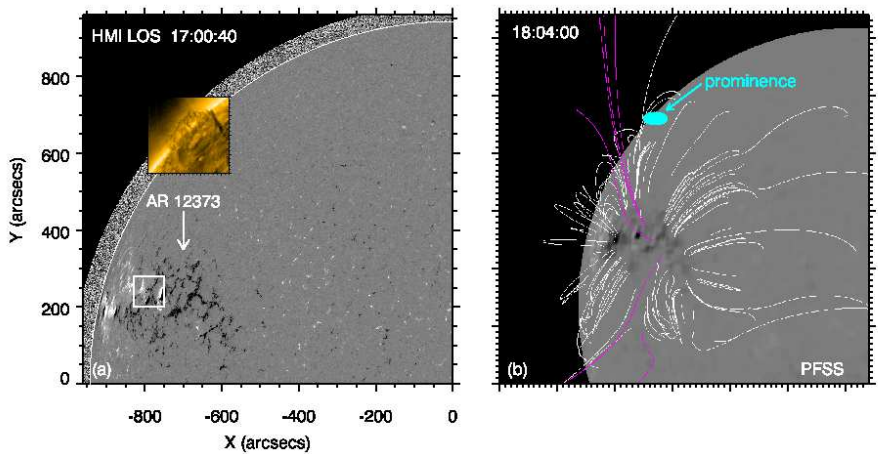}
\caption{\textit{Left panel:} HMI LOS magnetogram at 17:00:40 UT on June 29. The white box marks the field-of-view ($80\arcsec\times80\arcsec$) of Figure~\ref{fig6}.
The built-in figure is a closeup of the prominence in 171 {\AA}.
\textit{Right panel:} Global magnetic configuration using PFSS modeling at 18:04:00 UT. The white and purple lines represent the closed and open magnetic field lines.
The cyan oval indicates the rough location of prominence.
\label{fig2}}
\end{figure*}

In Figure~\ref{fig2}, the HMI LOS magnetogram at 17:00:40 UT is displayed in the left panel. The inset figure shows the prominence at 17:00:11 UT in 171 {\AA}. It is obvious that the photospheric magnetic field strength 
beneath the prominence is quite weak. The right panel demonstrates the global magnetic configuration at 18:04 UT using PFSS modeling. It is revealed that the AR and neighborhood of prominence are connected 
by closed magnetic field lines.

\subsection{C2.4 flare and coronal jet} \label{s-flare}
\begin{figure*}
\plotone{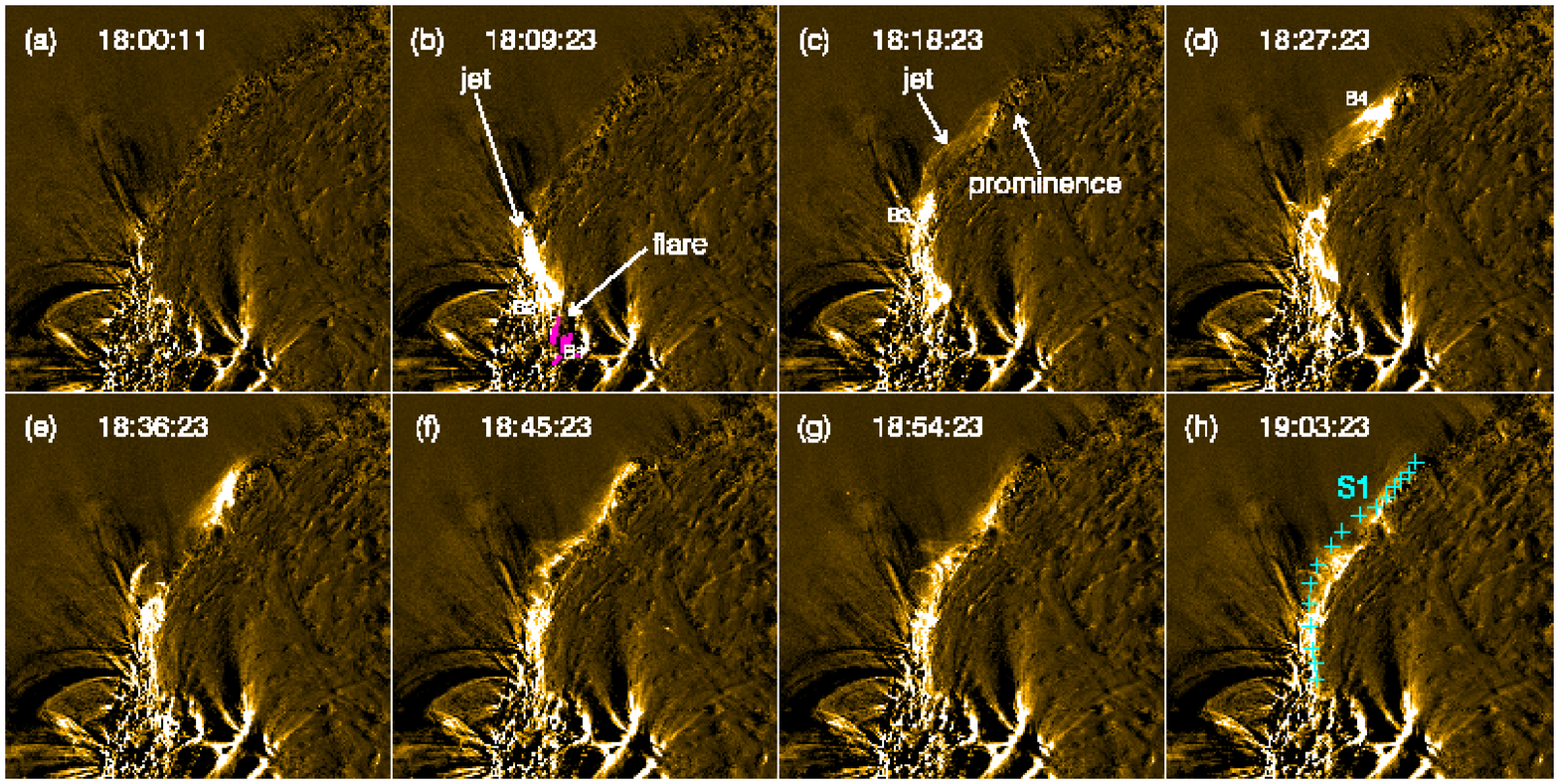}
\caption{Eight snapshots of the base-difference images in 171 {\AA} during 18:00$-$19:03 UT on June 29. The white arrows in panels (b)-(c) point at the flare, jet, and prominence.
In panel (b), the UV 1600 {\AA} intensity contours at 18:09:28 UT are superposed with thin magenta lines. In panel (h), the cyan plus symbols denote the curved slice S1 whose time-slice
diagrams are plotted in Figure~\ref{fig7}. (An animation of this figure is available.)
\label{fig3}}
\end{figure*}

During 17:58$-$19:00 UT, a C2.4 flare took place in AR 12373. To illustrate the flare more clearly, we took the EUV images around 17:00 UT as base images and made base-difference images after 17:00 UT. 
Figure~\ref{fig3} shows eight
snapshots of the base-difference images in 171 {\AA} (see also the online animated figure). At the very beginning of flare, there was no obvious brightening (see panel (a)). As time went on, the flare started to brighten.
The jet spurted out of the flare site and propagated northward (see panel (b)). About nine minutes later, part of the jet generated brightening at B3, while the remaining part propagated continuously along closed
coronal loops (see panel (c)). The jet stopped to the left of prominence and caused strong brightening at B4 (see panel (d)). Afterwards, the intensities of flare and jet decreased gradually.

\begin{figure*}
\plotone{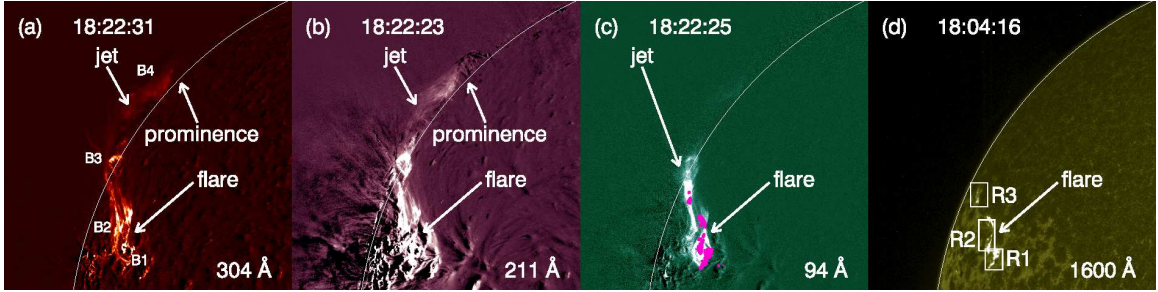}
\caption{(a-c) EUV base-difference images around 18:22:25 UT on June 29. (d) Original AIA 1600 {\AA} image at 18:04:16 UT. The field-of-view is $700\arcsec\times700\arcsec$.
The intensity contours of the three ribbons in panel (d) are superposed on the 94 {\AA} image with thin magenta lines.
\label{fig4}}
\end{figure*}

In Figure~\ref{fig4}, the EUV base-difference images around 18:22 UT are displayed in panels (a-c). The jet indicated in 171 {\AA} images are also visible in 304, 211, and 94 {\AA}, implying its 
multithermal nature \citep{zqm14}. The jet originated from B2 and went through B3 before terminating at B4 where it encountered the prominence. It is noticed that the jet connecting B3 and B4 is not so clear in 94 {\AA},
meaning that the temperature of that segment is not high enough ($<$6 MK).
Interestingly, the flare had three ribbons. Figure~\ref{fig4}(d) shows the original UV 1600 {\AA} image when the intensities of flare ribbons reached the maxima at 18:04:16 UT. 
The intensity contours of the three ribbons (R1$-$R3) are superposed on the 94 {\AA} image with thin magenta lines. R1 and R2 at the primary flare site are connected
by hot and compact post-flare loops. R3 is close to B3, suggesting that the primary flare site and remote site (R3) are connected by closed magnetic field lines.

\begin{figure}
\plotone{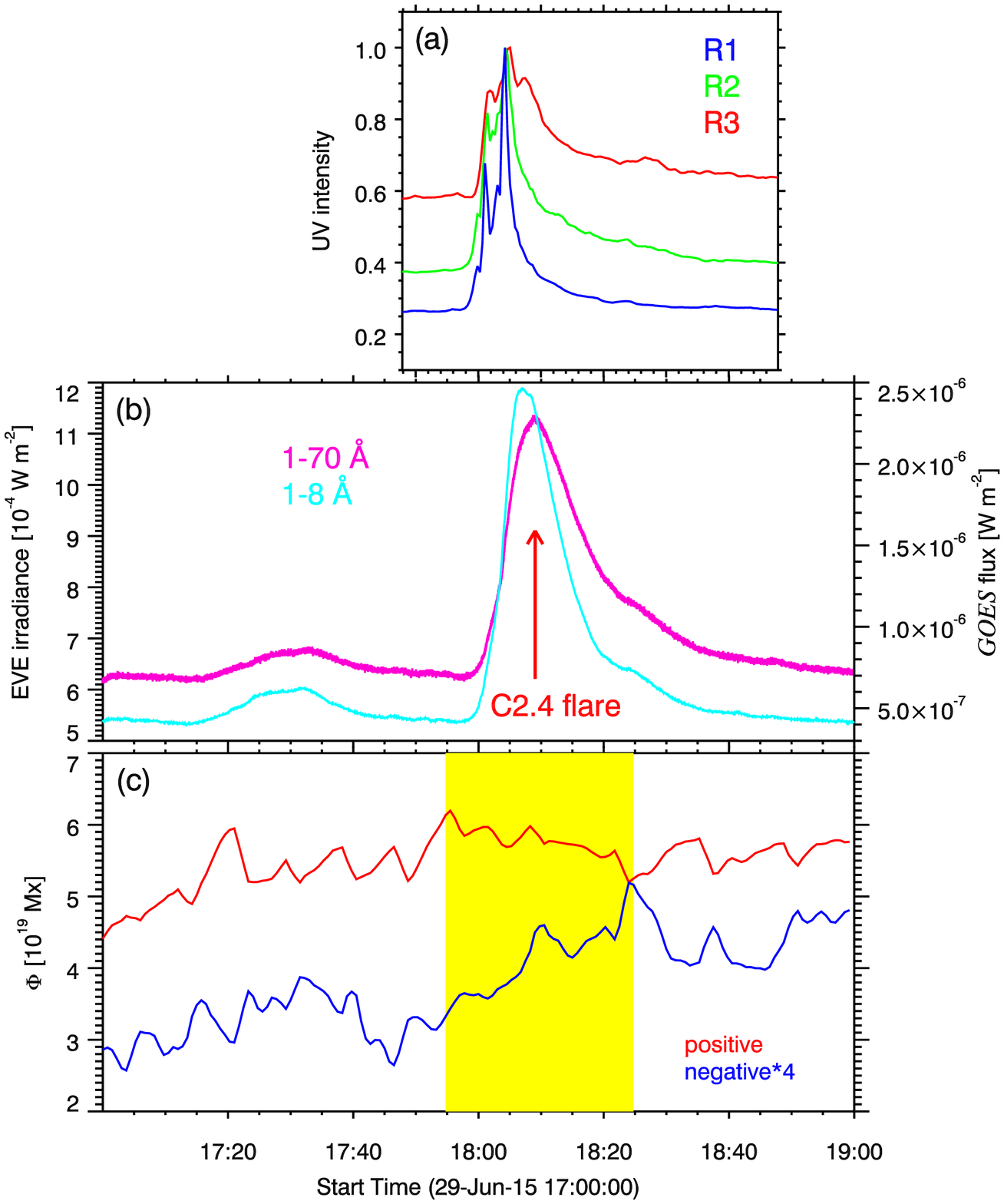}
\caption{(a) Temporal evolutions of the normalized UV intensities of the three ribbons (R1, R2, and R3) within the boxes of Figure~\ref{fig4}(d).
(b) Temporal evolutions of the EVE irradiance in 1$-$70 {\AA} (\textit{magenta line}) and \textit{GOES} SXR flux in 1$-$8 {\AA} (\textit{cyan line}).
The red arrow points at the peak time of flare. (c) Temporal evolutions of the unsigned positive (red line) and negative (blue line) magnetic 
fluxes within the black box of Figure~\ref{fig6}(e). The yellow rectangular region signifies the time of magnetic cancellation.
\label{fig5}}
\end{figure}

In Figure~\ref{fig5}(b), the SXR flux in 1$-$8 {\AA} and the EVE irradiance in 1$-$70 {\AA} during 17:00$-$19:00 UT are plotted with cyan and magenta lines, respectively.
The SXR flux increases sharply from $\sim$17:58 and reaches the peak value at $\sim$18:07 UT. Afterwards, the flux decreases rapidly to the initial level at $\sim$19:00 UT. The lifetime of flare is $\sim$1 hr.
The EVE irradiance has a similar evolution to the SXR flux except for a delayed peak time by $\sim$2 minutes. The temporal evolutions of the normalized UV intensities of R1$-$R3 are plotted in 
Figure~\ref{fig5}(a). There are two major peaks at $\sim$18:01 UT and $\sim$18:04 UT in the light curves. A close inspection reveals that the peak times of the ribbons are sequential rather than coincident.
Cross correlation analysis shows that the intensity of R3 lags behind that of R1 by $\sim$48 s. The free magnetic energy of the flare is converted into kinetic energy of jet, 
thermal/nonthermal energies of the plasmas, and MHD waves. So, the time lag between R1 and R3 implies that the nonthermal energy is transported from the primary flare site to remote site by the
high-energy (10$-$100 keV) electrons \citep{naka85}.

\begin{figure*}
\plotone{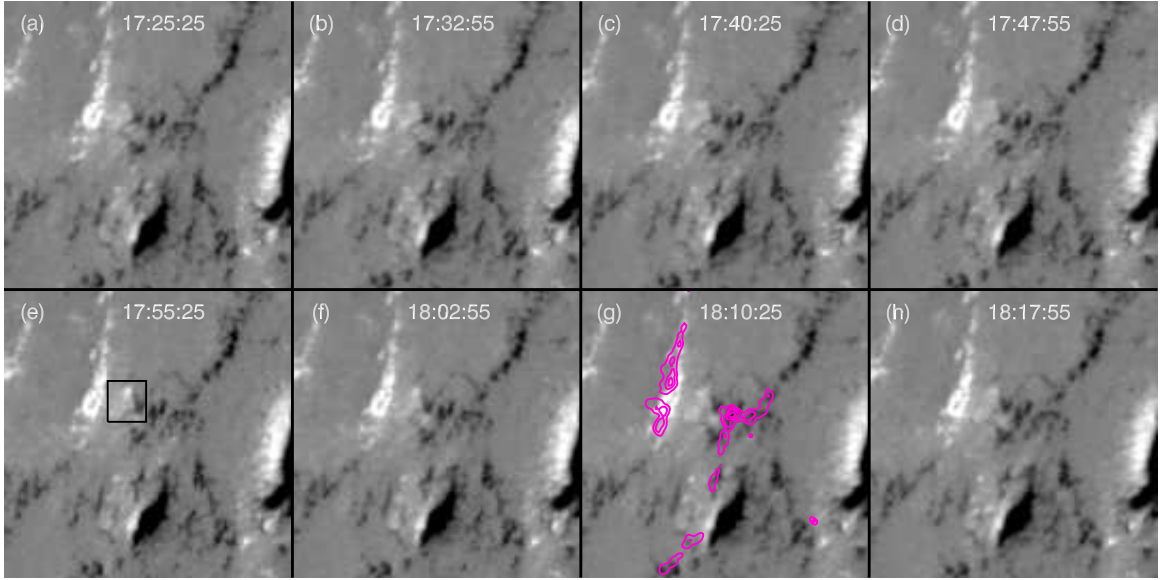}
\caption{Eight snapshots of the HMI LOS magnetograms with a field-of-view of $80\arcsec\times80\arcsec$ during 17:25$-$18:18 UT.
White/black colors represent positive/negative magnetic polarities with gray scale levels of $\pm$400 G.
In panel (e), a black box marks the region where magnetic cancelation took place.
In panel (g), the UV 1600 {\AA} intensity contours at 18:09:28 UT are superposed with thin magenta lines.
\label{fig6}}
\end{figure*}

It is generally believed that coronal jets are driven by magnetic reconnection \citep[e.g.,][]{yoko96,more08,arch13,ni17}. In order to figure out whether the flare-related jet is associated with magnetic reconnection in this event, 
we examined the HMI LOS magnetograms with a cadence of 45 s. It is found that there was magnetic cancellation at the jet base. Eight snapshots of the magnetograms during 17:25$-$18:18 UT are displayed in Figure~\ref{fig6}. 
In panel (e), the region of magnetic cancellation is marked by the black box. The unsigned integrated positive and negative magnetic fluxes within the box are calculated and plotted with red and blue lines in Figure~\ref{fig5}(c).
It is obvious that the positive magnetic flux increased from 4.5$\times10^{19}$ Mx at 17:00 UT to 6.2$\times10^{19}$ Mx at 17:55 UT. During the impulsive and decay phases of the flare, the positive magnetic flux was continuously 
cancelled by the negative field (see the yellow rectangular region). The negative flux increased slightly during the cancellation, probably due to that the rate of flux emergence exceeds the rate of cancellation.
Therefore, the flare-related coronal jet may be driven by magnetic reconnection as indicated by flux cancellation near the flaring site.

\subsection{Prominence oscillations} \label{s-osci}

\subsubsection{Transverse oscillation of EP} \label{s-ep}
\begin{figure*}
\plotone{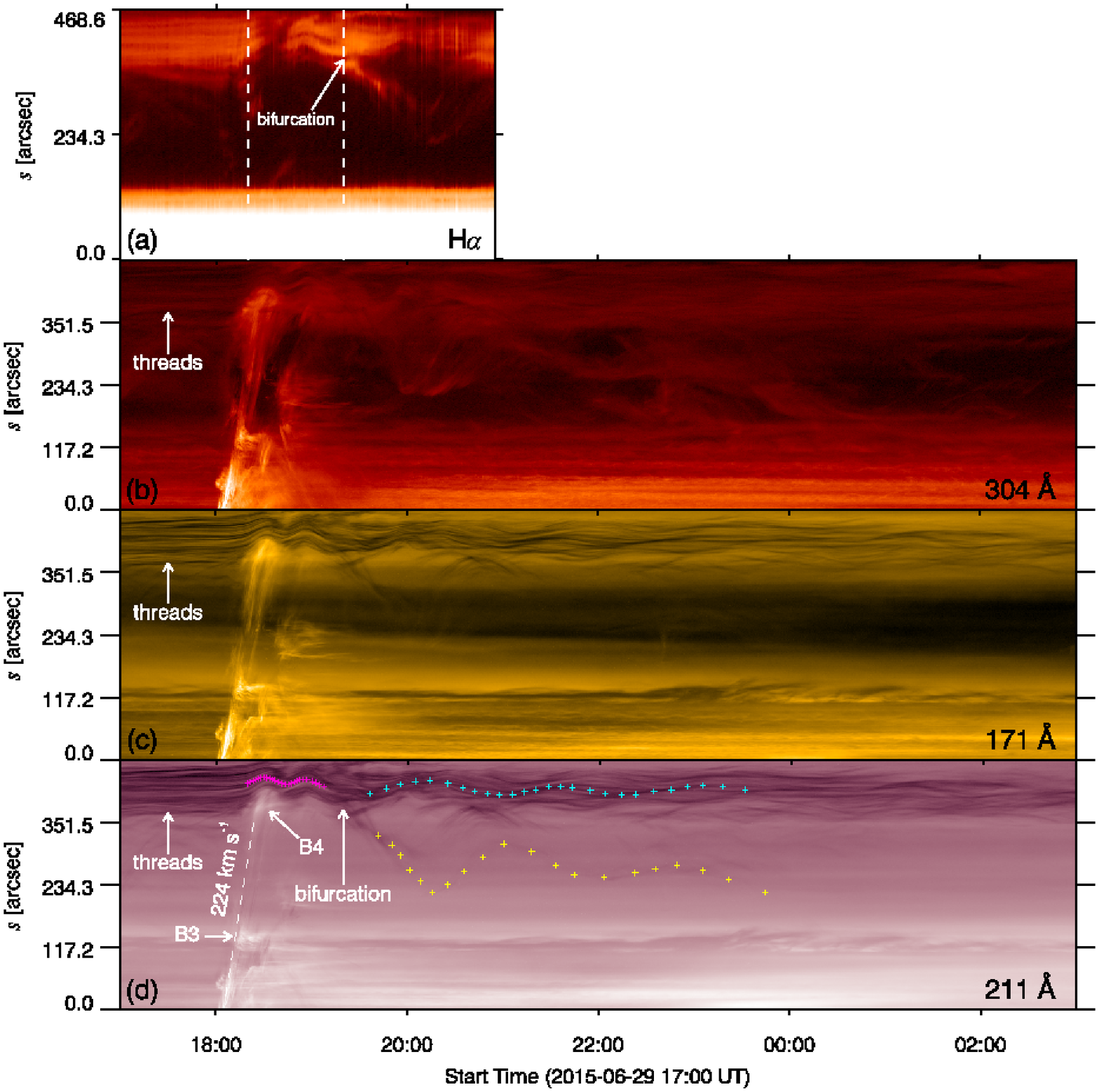}
\caption{Time-slice diagrams of S1 in various wavelengths. $s=0$ and $s=468\farcs$6 in the $y$-axis stand for the southeast and northwest endpoints of S1. 
In panel (a), the two vertical dashed lines mark the times during which the prominence oscillation is obvious in H$\alpha$. The arrow indicates the bifurcation point of oscillations.
In panels (b-d), the vertical arrows point at the dark threads of EP. 
In panel (d), the apparent velocity of the coronal jet is labeled. The short arrows point at the rough positions of B3 and B4. 
The magenta and cyan plus symbols are manually marked positions of vertical threads, which are used for curve fitting in Figure~\ref{fig8}.
The yellow plus symbols are manually marked positions of the horizontal threads, which are used for curve fitting in Figure~\ref{fig12}.
\label{fig7}}
\end{figure*}

From the online animation of Figure~\ref{fig3}, we found that the jet was stopped by the prominence. However, due to the strong impact of jet, the prominence started to oscillate periodically.
In order to investigate the jet and prominence oscillations, we take a long curved slice (S1) with a length of 468$\farcs$6 in Figure~\ref{fig3}(h). The slice starts from the flare site and goes through the jet and 
EP of prominence. The time-slice diagrams of S1 in various wavelengths are demonstrated in Figure~\ref{fig7}.
It is clear that the jet propagated from the primary flare site to the remote site (B3) during 18:04$-$18:16 UT. The constant apparent velocity ($\sim$224 km s$^{-1}$) is the same in various EUV wavelengths. 
The velocity of jet is comparable to the velocities of jets propagating along a closed magnetic loop and generating sympathetic coronal bright points \citep{zqm16}.
The remaining part of jet continues to propagate forward along S1 before terminating to 
the east of prominence around 18:30 UT, which is accompanied by strong brightening (B4). The prominence was impulsively pushed aside, moving in the northwest direction. The restoring force makes the 
prominence decelerate and then move in the opposite direction. Such a cyclic motion, i.e., oscillation, continues for several hours, which is clearly demonstrated in Figure~\ref{fig7}. Like most of the cases 
reported in previous literatures, the oscillation is damping. In other words, the amplitude attenuates with time and disappears after several cycles. Owing to the lower resolution 
and cadence of H$\alpha$ observation, the oscillation in H$\alpha$ is not as obvious as that in the EUV wavelengths. However, the oscillation during 18:20$-$19:20 UT is identifiable and in phase with the EUV wavelengths.

\begin{figure}
\plotone{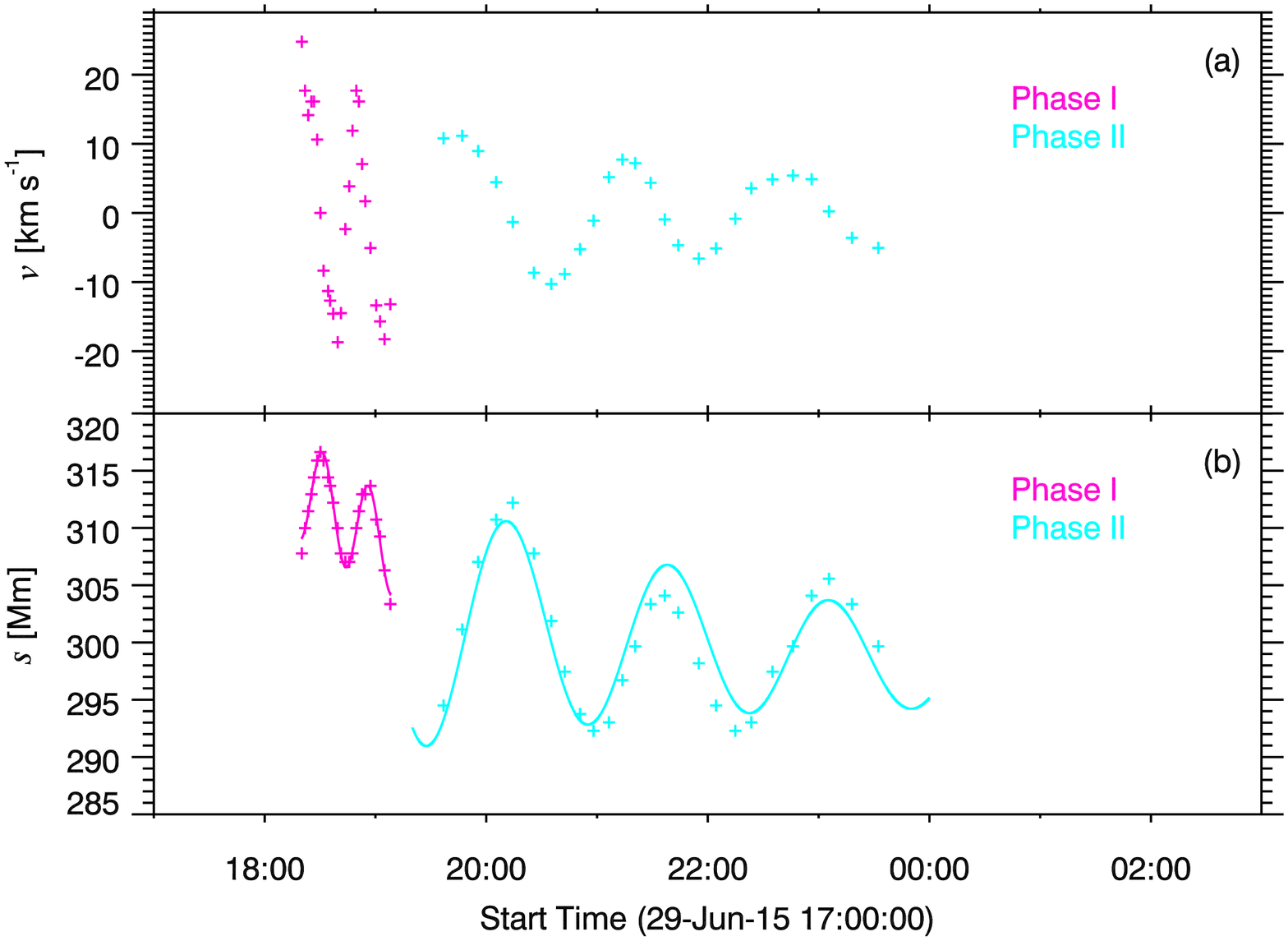}
\caption{(a) Derived velocity of the oscillating prominence EP along S1. 
(b) Positions of the prominence EP along S1 (plus symbols) and the fitted curves (solid lines).
The magenta and cyan colors stand for the results of phase I and phase II, respectively.
\label{fig8}}
\end{figure}

In order to precisely calculate the parameters of prominence oscillation, we mark the positions of EP manually with magenta and cyan plus symbols in Figure~\ref{fig7}(d). Then, we fit the curves by using the standard
program \textit{mpfit.pro} in \textit{SSW} and the function \citep{zqm17}
\begin{equation} \label{eqn-1}
y=y_0+bt+A_0\sin(\frac{2\pi}{P}t+\phi_0)e^{-t/\tau},
\end{equation}
where $y_0$, $A_0$, and $\phi_0$ represent the initial position, amplitude, and phase. $b$, $P$, and $\tau$ stand for the linear velocity of the threads, period, and damping timescale of the oscillation.
Since the periods of oscillation are fairly different before and after $\sim$19:20 UT, we divide the evolution into two phases, phase I (18:20$-$19:20 UT) and phase II (19:20$-$23:59 UT). In Figure~\ref{fig8}, we plot the
results of curve fitting. The parameters are also listed in the top two rows of Table~\ref{tab:fitting}. In phase I, the amplitude and period of oscillation are $\sim$4.5 Mm and $\sim$25 minutes. The maximal velocity of 
oscillation can reach $\sim$20 km s$^{-1}$. The oscillation lasts for about two cycles with a damping timescale of $\sim$7.5 hr. Hence, the damping ratio ($\tau/P$) is calculated to be $\sim$18.2. In phase II, 
the amplitude of oscillation increases and is $\sim$2.5 times larger than that in phase I. The period also increases and is $\sim$3.5 times larger than that in phase I, implying that the restoring force decreases as time goes by.
In addition, the maximal velocity of oscillation halves to $\sim$10 km s$^{-1}$. The oscillation lasts for at least three cycles with a damping timescale of $\sim$4.4 hr. Hence, the damping ratio is calculated to be $\sim$3,
suggesting that the attenuation of oscillation in phase II becomes faster. It should be emphasized that the oscillations of the fine threads of prominence are very complicated and vary from point to point (see Figure~\ref{fig7}), 
we track the positions of the darkest thread, which is representative of EP.

\begin{deluxetable*}{cccccccc}
\tablecaption{Fitted parameters of the prominence oscillations \label{tab:fitting}}
\tablecolumns{8}
\tablenum{2}
\tablewidth{0pt}
\tablehead{
\colhead{Time} &
\colhead{$b$} &
\colhead{$A_0$} &
\colhead{$\phi_0$} &
\colhead{$P$} &
\colhead{$\tau$} &
\colhead{$\tau/P$} &
\colhead{type}\\
\colhead{(UT)} & 
\colhead{(km s$^{-1}$)} &
\colhead{(Mm)} &
\colhead{(rad)} &
\colhead{(min)} &
\colhead{(hr)} &
\colhead{} &
\colhead{}
}
\startdata
18:20$-$19:20\tablenotemark{a} & -1.80 &  4.49 & -1.30 & 24.85 & 7.53 & 18.18 & transverse\\
19:20$-$23:59\tablenotemark{b} & -0.23 & 11.27 & -2.15 & 87.41 & 4.42 & 3.03 & transverse\\
18:20$-$20:20\tablenotemark{c} &  0.68 &  2.03 & -1.92 & 26.87 & 1.74 & 3.89 & transverse\\
19:20$-$23:59\tablenotemark{d} & -1.86 & 52.44 &  0.83 & 98.72 & 2.50 & 1.52 & longitudinal\\
\enddata
\tablenotetext{a}{Phase I of the prominence oscillation of EP}
\tablenotetext{b}{Phase II of the prominence oscillation of EP}
\tablenotetext{c}{Time of the prominence oscillation of WP}
\tablenotetext{d}{Time of the horizontal threads oscillation}
\end{deluxetable*}

\subsubsection{Transverse oscillation of WP} \label{s-wp}
\begin{figure}
\plotone{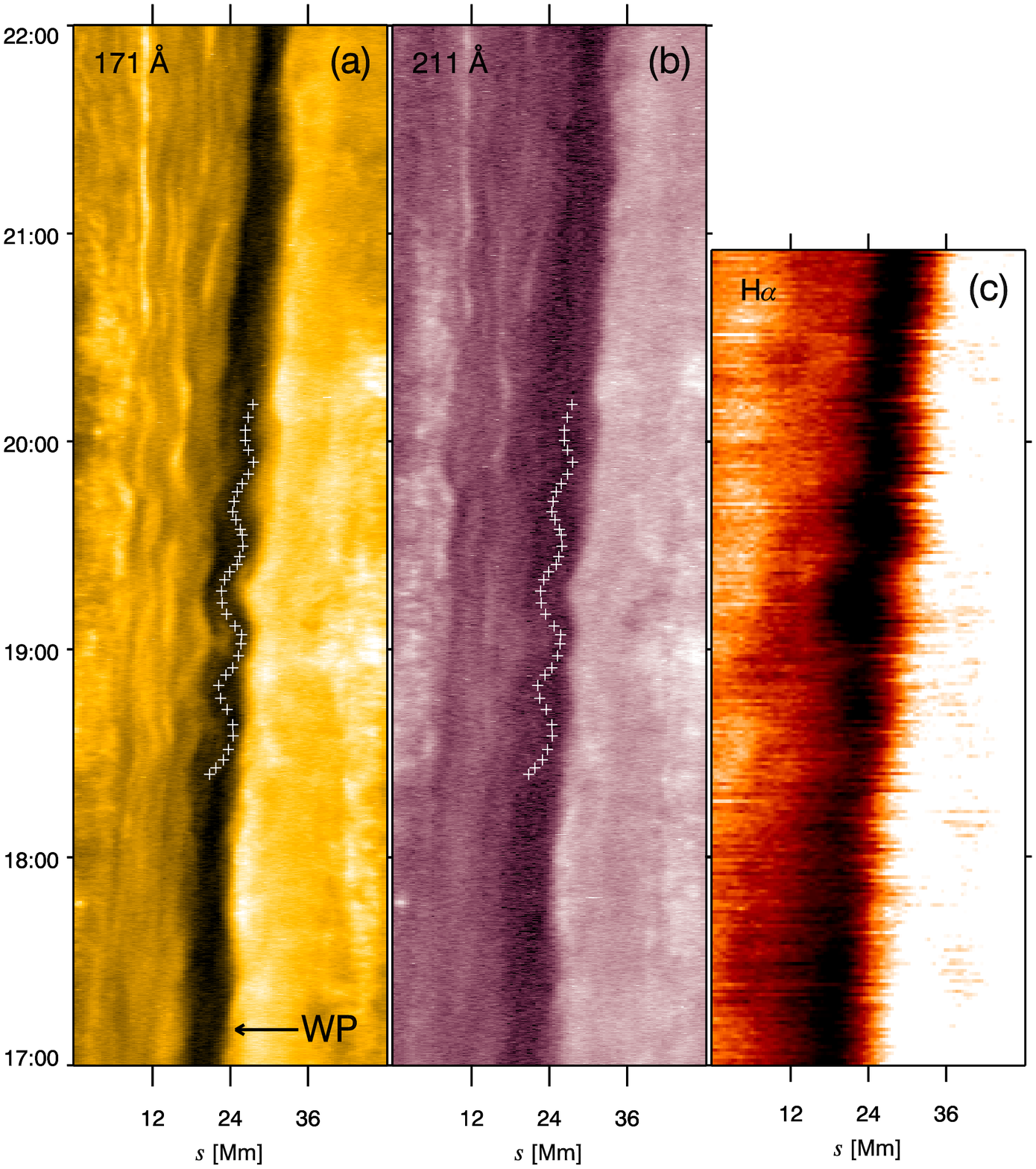}
\caption{Time-slice diagrams of S2 during 17:00$-$22:00 UT in 171 {\AA} (a), 211 {\AA} (b), and H$\alpha$ (c).
The $y$-axis is UT time on June 29. $s=0$ and $s=48$ Mm in the $x$-axis denote the southeast and northwest endpoints of S2 (see Figure~\ref{fig1}(a)).
The white plus symbols are manually marked positions of WP.
\label{fig9}}
\end{figure}

Like the EP of prominence, the WP of prominence also experienced transverse oscillation. In Figure~\ref{fig1}(a), we extract the intensity along a second slice (S2), which has a length of 
$\sim$48 Mm and is perpendicular to WP. The time-slice diagrams of S2 in 171 {\AA}, 211 {\AA}, and H$\alpha$ are demonstrated in Figure~\ref{fig9}. 
$s=0$ and $s=48$ Mm in the $x$-axis denote the southeast and northwest endpoints of S2. It is obvious that the dark WP underwent transverse oscillation during 18:20$-$20:20 UT in 
EUV and H$\alpha$ wavelengths. The initial direction of movement of WP is consistent with that of EP, meaning that the whole prominence oscillates coherently.
Although the resolution and cadence of H$\alpha$ observation are lower than AIA, we can still find that the oscillation in H$\alpha$ and EUV wavelengths are completely in phase.

\begin{figure}
\plotone{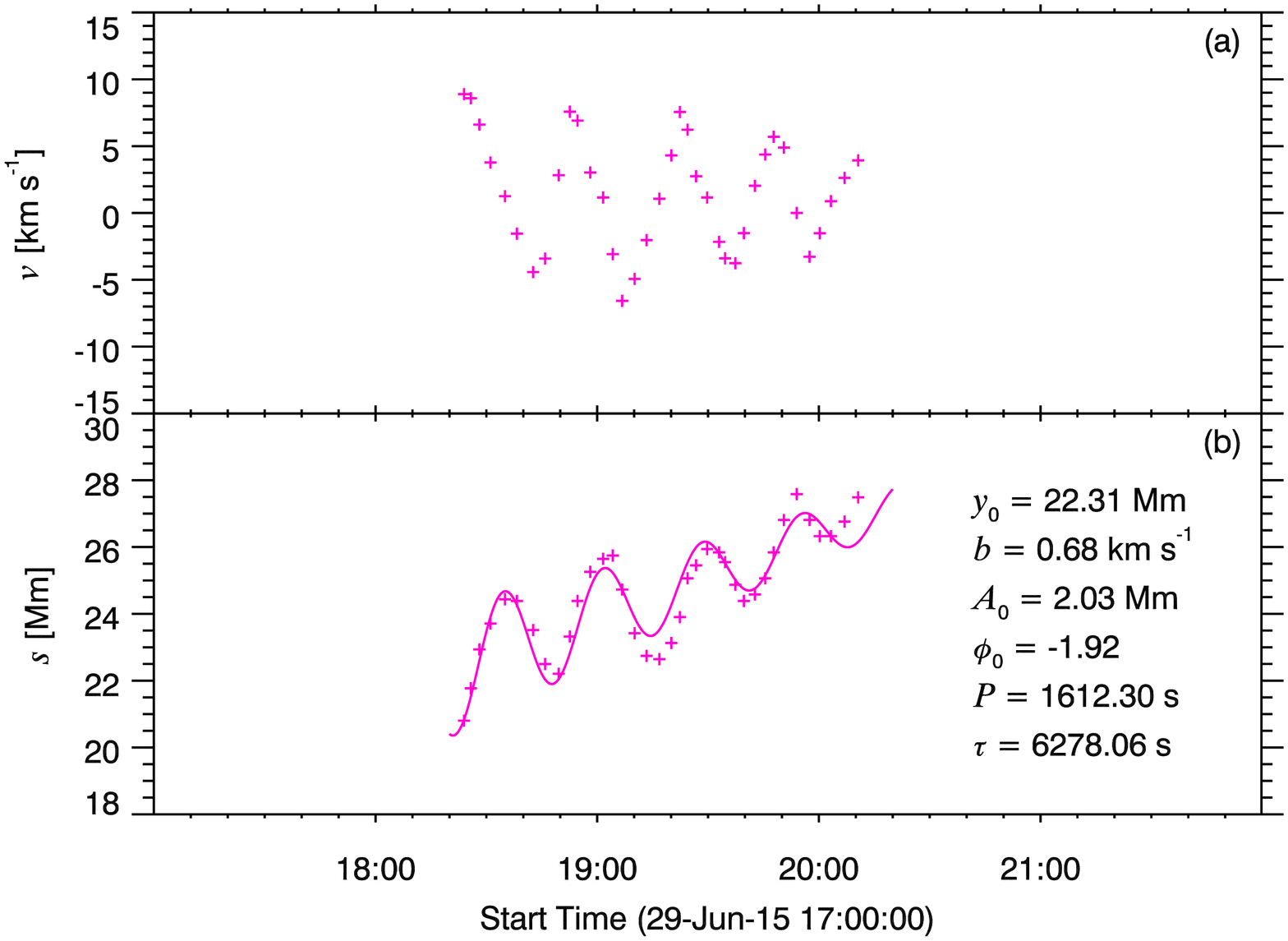}
\caption{(a) Derived velocity of the oscillating prominence WP along S2. (b) Positions of prominence WP along S2 (plus symbols) and the fitted curve (solid line).
\label{fig10}}
\end{figure}

Like in Figure~\ref{fig7}(d), we mark the positions of WP with white plus symbols in Figure~\ref{fig9} and perform curve fitting. In Figure~\ref{fig10}, we plot the results of curve fitting using Equation~\ref{eqn-1}.
The parameters are listed in the third row of Table~\ref{tab:fitting}. Compared with EP, both the amplitude ($\sim$2 Mm) and maximal velocity ($<$10 km s$^{-1}$) of WP are much smaller. 
The period ($\sim$27 minutes) is slightly longer than that of EP in phase I. However, the attenuation of the transverse oscillation of WP is the fastest 
with a timescale of 1.74 hr and a damping ratio of $\sim$3.9, which can explain why the transverse oscillation of WP lasts for only $\sim$4 cycles until 20:20 UT.

\subsubsection{Longitudinal oscillation of the horizontal threads} \label{s-ht}
\begin{figure*}
\plotone{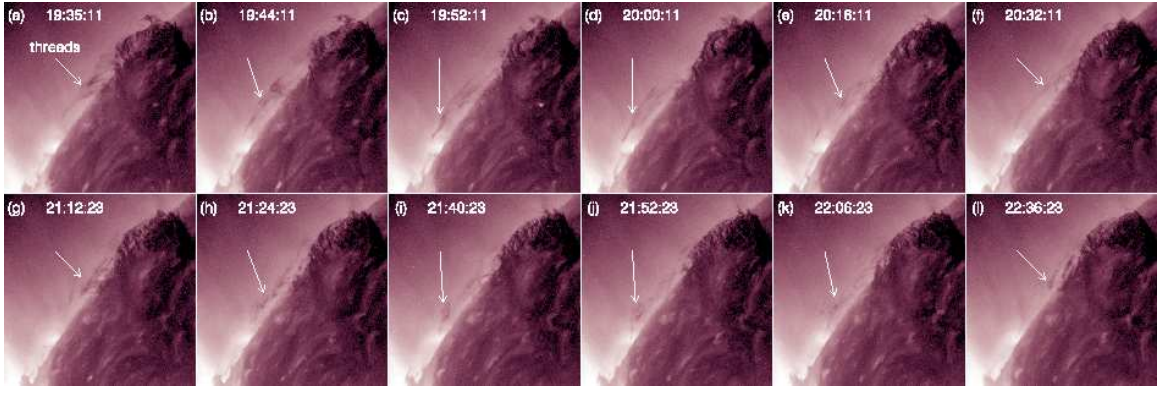}
\caption{Twelve snapshots of AIA EUV images in 211 {\AA}. The arrows point at the horizontal threads oscillating along the arcade to the east of EP. 
(An animation of this figure is available.)
\label{fig11}}
\end{figure*}

Apart from the transverse oscillations of the prominence, a handful of horizontal threads oscillated along the arcade to the east of EP (see also the online animated figure). 
Figure~\ref{fig11} shows 12 snapshots of AIA EUV images in 211 {\AA}. From $\sim$19:35, the dark threads moved in the southeast direction until $\sim$20:05 UT. Afterwards, the 
threads moved reversely, i.e., in the northwest direction until $\sim$21:00 UT. Then, a second cycle of oscillation of the threads took place during 21:00$-$22:40 UT (see the bottom panels). 
After careful inspection of the online animation, a total of 2.5 cycles can be identified. Such longitudinal oscillation of the horizontal threads is similar to the case of AR prominence \citep{zqm12}.
It is also evident in the time-slice diagrams of S1, especially in 304 {\AA}, since S1 goes through the long arcade.  

\begin{figure}
\plotone{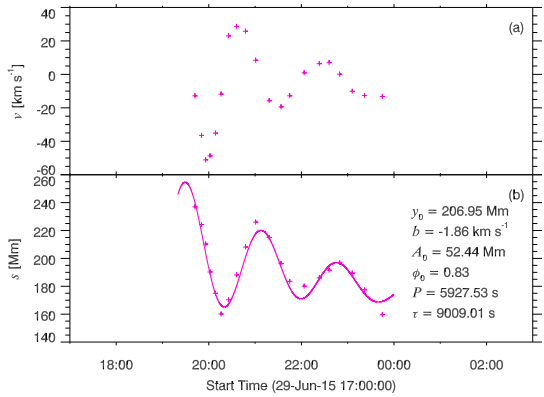}
\caption{(a) Derived velocity of the horizontal threads along S1. 
(b) Positions of the horizontal threads along S1 (plus symbols) and the fitted curves (solid lines).
\label{fig12}}
\end{figure}

In Figure~\ref{fig7}(d), we mark the positions of horizontal threads with yellow plus symbols. In order to calculate the parameters of oscillation, we perform curve fitting. In Figure~\ref{fig12}, we plot the results of curve 
fitting using Equation~\ref{eqn-1}. The parameters are listed in the fourth row of Table~\ref{tab:fitting}. Compared with the transverse oscillations of prominence, the initial amplitude ($\sim$52.4 Mm), 
period ($\sim$99 minutes), and peak velocity ($\sim$50 km s$^{-1}$) of the horizontal threads oscillation are remarkably larger. The damping timescale is 2.5 hr and the damping ratio is $\sim$1.5, suggesting a faster attenuation.

\section{Discussion} \label{sec:discuss}

\subsection{How are the prominence oscillations triggered?} \label{sec:trigger}
The triggering mechanism of large-amplitude prominence oscillations is an important issue. In most cases of longitudinal oscillations, they are triggered by subflares or microflares at the footpoints of filaments 
\citep[e.g.,][]{jing03,jing06,vrs07,li12}. The local brightenings at the footpoints are sometimes associated with intermittent jets that propagate upward and drive filament oscillations \citep{luna14}. 
The magnetic reconnections during subflares or microflares result in impulsive heating at the chromosphere so that the gas pressure is greatly enhanced, pushing the filament material to oscillate \citep{zqm13,zhou17}. 
Occasionally, when an incoming shock wave from a remote flare encounters a filament, it is likely that it triggers longitudinal or transverse filament oscillations depending on the incident direction \citep{shen14}. 
For transverse prominence oscillations, most of them are triggered by large-scale EUV or Moreton waves from a remote site of eruption \citep{ram66,eto02,oka04,gil08,her11,dai12,gos12}, though a few of them 
are associated with emerging fluxes \citep{chen08}. 

\begin{figure*}
\plotone{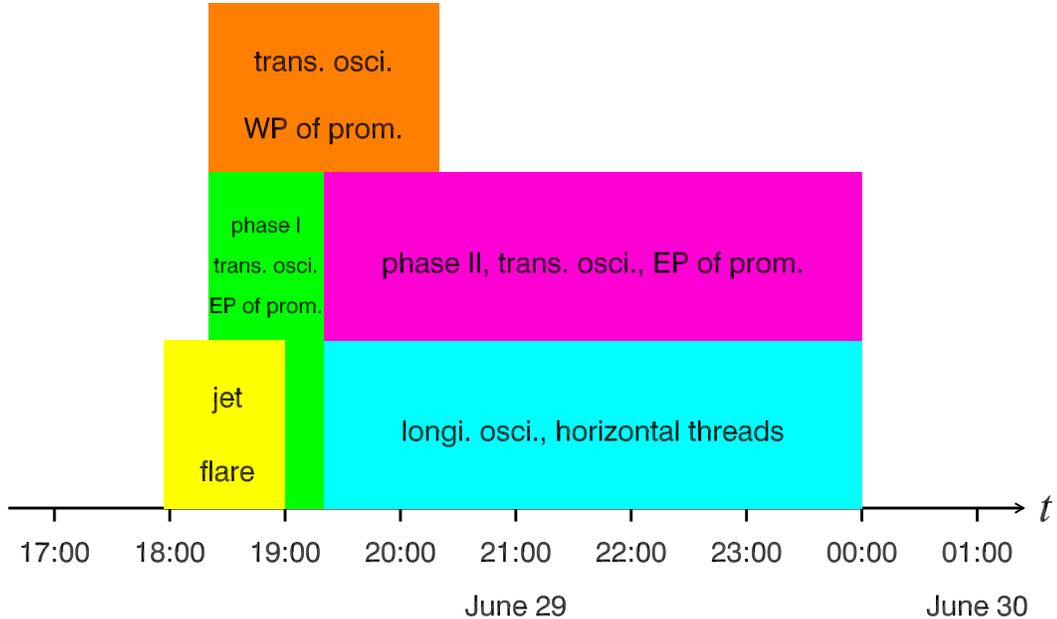}
\caption{Timeline of all phenomena, including the C-class flare, coronal jet, transverse oscillations of the EP and WP of prominence, and longitudinal oscillation of the horizontal threads.
\label{fig13}}
\end{figure*}

In this study, simultaneous transverse and longitudinal oscillations are triggered by a coronal jet from the remote C2.4 flare in AR 12373, which has never been noticed before.
The AR and quiescent prominence, which has a long distance of $\sim$255 Mm, are connected by closed magnetic field lines, so that the jet from the primary flare site can reach and interact with the prominence.
As is described in Section~\ref{s-osci}, the oscillations of the prominence and horizontal threads are very complex. For the transverse oscillation of the EP of prominence, the parameters of the two phases, 
including the amplitudes, velocities, periods, and damping times, are totally different. A question is raised: What is the cause of big difference? From Figure~\ref{fig7} and the online movie of Figure~\ref{fig11}, we 
noticed that the longitudinal oscillation of the horizontal threads was coincident with phase II of the oscillation of EP. Meanwhile, the material of horizontal threads came from EP. 
Therefore, we can draw a conclusion that there was a bifurcation at the end of phase I of the oscillation of EP (see Figure~\ref{fig7}(d)). The material escaping from EP to the long arcade became the horizontal 
threads that underwent longitudinal oscillation. The remaining material of EP continued oscillating in phase II. However, the transfer of mass may change the magnetic configuration of EP, resulting in fairly different 
parameters of oscillation in phase II. From Figure~\ref{fig7} and Figure~\ref{fig9}, we found that the onsets of transverse oscillations of WP and EP were coincident. Moreover, the periods of WP and EP during phase I 
are very close, indicating that the prominence oscillated coherently. Since WP is far from the horizontal threads, the transverse oscillation of WP was not disturbed or disrupted by the mass transfer at the end of phase I.
The timeline of all phenomena are illustrated in Figure~\ref{fig13}.

It should be emphasized that precise identification of the mode of oscillations suffers from the lack of spectroscopic observation and stereoscopic observation from two perspectives. 
On one hand, the 3D morphology of prominence is unclear from a single perspective. On the other hand, the Doppler velocity of the prominence is unavailable. We are not sure whether the prominence and 
horizontal threads oscillate in the LOS direction. Additional case studies using spectroscopic and stereoscopic observation are worthwhile to investigate the large-amplitude prominence oscillations.

\subsection{Curvature radius and magnetic field strength of the dip} \label{sec:radius}
Magnetic tension force is widely accepted to be the restoring force of transverse prominence oscillations  \citep{kle69}. However, the restoring force of longitudinal prominence oscillations remain unclear hitherto. 
In the context of one-dimensional (1D) model, gravity of the prominence along the magnetic dip is considered to be the dominant restoring force \citep{luna12a,zqm12,zqm13,luna14,luna16a,luna16b}, although the gas 
pressure gradient is not neglectable for very shallow dips. In this study, the longitudinal oscillation of the horizontal threads along the arcade can easily been understood using the pendulum model. 
The period is expressed as $P=2\pi \sqrt{R/g_{\odot}}$, where $R$ denotes the curvature radius of the dip and $g_{\odot}$ represents the gravitational acceleration at the solar surface.
Therefore, the curvature radius of the long arcade supporting the threads is estimated to be $\sim$244 Mm. In addition, we can estimate the lower limit of the transverse magnetic field strength of the arcade
using a simple analytical expression ($B_{tr}$[G]$\geq$17$P$[hr]) \citep{luna14}. Taking the observed value of $P$ (1.65 hr), one can derive the lower limit of $B_{tr}$$\geq$28 G.

\section{Summary} \label{sec:summary}
In this paper, we report our multiwavelength observations of a quiescent prominence observed by GONG and \textit{SDO}/AIA on 2015 June 29. The main results are summarized as follows:
\begin{enumerate}
\item{A C2.4 flare occurred with a lifetime of $\sim$1 hr in AR 12373, which was associated with a pair of short ribbons and a remote ribbon in the chromosphere.
During the impulsive phase of the flare, a coronal jet spurted out of the primary flare site and propagated in the northwest direction at an apparent speed of $\sim$224 km s$^{-1}$.
Part of the jet stopped near the remote ribbon and generated brightenings in various EUV wavelengths. The remaining part, however, continued moving and terminated 
to the east of the prominence.}
\item{Once the jet encountered the prominence, it produced localized brightening and pushed the prominence to oscillate periodically. The EP of prominence, consisting of many vertical threads,
experienced transverse oscillation, which can be divided into two phases. In phase I, the initial amplitude, velocity, and period are $\sim$4.5 Mm, $\sim$20 km s$^{-1}$, and $\sim$25 minutes, respectively. 
The oscillation lasted for about two cycles with a damping timescale of $\sim$7.5 hr. In phase II, the initial amplitude increases to $\sim$11.3 Mm, which is $\sim$2.5 times larger than that of the first phase.
The initial velocity, however, halves to $\sim$10 km s$^{-1}$. The period increases by a factor of $\sim$3.5, indicating that the restoring force reduced in phase II.
The oscillation lasted for about three cycles, with the damping timescale decreasing significantly, which means that the attenuation of the oscillation became faster.}
\item{The WP of prominence also underwent transverse oscillation. The initial amplitude is only $\sim$2 Mm and the velocity is $<$10 km s$^{-1}$.
The period ($\sim$27 minutes) is slightly longer than that of EP in phase I. The oscillation lasted for about four cycles with the shortest damping timescale ($\sim$1.7 hr).}
\item{To the east of the prominence, a handful of horizontal threads experienced longitudinal oscillation along an arcade. The initial amplitude, velocity, 
and period are $\sim$52.4 Mm, $\sim$50 km s$^{-1}$, and $\sim$99 minutes, respectively. The oscillation lasted for $\sim$2.5 cycles with a damping timescale of $\sim$2.5 hr.
The oscillation of the horizontal threads can be explained by the 1D pendulum model where projected gravity of the threads serves as the restoring force.
The curvature radius ($\sim$244 Mm) and the lower limit of magnetic field strength ($\sim$28 G) of the arcade are estimated.
Additional case studies and numerical simulations are required to investigate large-amplitude prominence oscillations.}
\end{enumerate}

\acknowledgments
We would like to thank H. S. Ji, Y. N. Su, V. Nakariakov, P. F. Chen, Y. Guo, Y. H. Zhou, J. T. Su, S. H. Yang, X. L. Yan, and Y. D. Shen for fruitful and valuable discussions. 
Q. M. Zhang acknowledges support from the International Space Science Institute (ISSI) to the Team 314 on ``Large-Amplitude Oscillation in prominences" led by M. Luna.
\textit{SDO} is a mission of NASA\rq{}s Living With a Star Program. AIA and HMI data are courtesy of the NASA/\textit{SDO} science teams. 
This work utilizes GONG data from NSO, which is operated by AURA under a cooperative agreement with NSF and with additional financial support from NOAA, NASA, and USAF.
This work is supported by the Youth Innovation Promotion Association CAS, NSFC (Nos. 11333009, 11773079, 11603077, 11573072), the Fund of Jiangsu Province (Nos. BK20161618 and BK20161095), 
CAS Key Laboratory of Solar Activity, National Astronomical Observatories (KLSA201716).

\end{document}